\documentclass[10pt]{iopart}
\usepackage{amssymb}
\usepackage{color}
\usepackage{graphicx}% Include figure files
\usepackage{csquotes}
\usepackage[pagewise, mathlines]{lineno}
\expandafter\let\csname equation*\endcsname\relax

\expandafter\let\csname endequation*\endcsname\relax

\usepackage{amsmath}
\newcommand*\patchAmsMathEnvironmentForLineno[1]{%
  \expandafter\let\csname old#1\expandafter\endcsname\csname #1\endcsname
  \expandafter\let\csname oldend#1\expandafter\endcsname\csname end#1\endcsname
  \renewenvironment{#1}%
     {\linenomath\csname old#1\endcsname}%
     {\csname oldend#1\endcsname\endlinenomath}}% 
\newcommand*\patchBothAmsMathEnvironmentsForLineno[1]{%
  \patchAmsMathEnvironmentForLineno{#1}%
  \patchAmsMathEnvironmentForLineno{#1*}}%
\AtBeginDocument{%
\patchBothAmsMathEnvironmentsForLineno{equation}%
\patchBothAmsMathEnvironmentsForLineno{align}%
\patchBothAmsMathEnvironmentsForLineno{flalign}%
\patchBothAmsMathEnvironmentsForLineno{alignat}%
\patchBothAmsMathEnvironmentsForLineno{gather}%
\patchBothAmsMathEnvironmentsForLineno{multline}%
}

\begin{document}

\title[Crystal growth in nano-confinement]
{Crystal growth in nano-confinement:\\
Subcritical cavity formation and viscosity effects}
\author{Luca Gagliardi and Olivier Pierre-Louis}
\address{CNRS, ILM Institut Lumi\`ere Mati\`ere,\\
Universit\' e Claude Bernard Lyon 1
Campus LyonTech-La Doua
Batiment Brillouin, 10 rue Ada Byron 
F-69622 Villeurbanne, France}
\eads{\mailto{luca.gagliardi@univ-lyon1.fr}, \mailto{olivier.pierre-louis@univ-lyon1.fr}}

\begin{abstract}
 We report on the modeling
of the formation of a cavity at the surface of crystals confined by a flat wall
during growth in solution.  
Using a continuum thin film model, we discuss two phenomena that could be observed when decreasing the 
thickness of the liquid film between the crystal and the wall down to the nanoscale. 
First, in the presence of an attractive van der Waals contribution to the disjoining
pressure, the formation of the cavity becomes sub-critical, i.e., discontinuous. 
In addition, there is a minimum supersaturation
required to form a cavity.
Second, when the thickness of the liquid film between
the crystal and the substrate reaches the nanoscale,
viscosity becomes relevant and hinders the formation of the cavity. 
We demonstrate that there is a critical value of the viscosity 
above which no cavity will form. 
The critical viscosity increases as the square of the 
thickness of the liquid film.
A quantitative discussion of model materials such as Calcite, 
Sodium Chlorate, Glucose and Sucrose is provided.
\end{abstract}


\maketitle

\section{Introduction}\label{sec:intro}

Crystal growth 
is commonly confined in pores, faults, or gaps,
as observed for example in rocks,
in natural and artificial cements, or in biomineralization. 
In these conditions, crystals can be directly formed
on substrate surfaces ---such as during heterogeneous nucleation~\cite{Markov2017,Turnbull1950,Winter2009,Page2006,Chayen2006}, 
or can be sedimented on substrates due to gravity.
The subsequent growth then occurs in the presence of
a contact with a substrate.
Here, we wish to discuss the growth dynamics
with the simplest type of contact, i.e.
with a flat, rigid, and impermeable wall. 

While growth can then occur at the free surface away
from the contacts via bulk transport of growth units,
growth in the contact regions requires mass transport 
along the interface between the crystal and the substrate~\cite{Durney1972}
when the substrate is impermeable.
The presence of a liquid film in the contact is a key ingredient to 
allow for such mass transport along the interface
during solution growth, as discussed
in the literature~\cite{Weyl1959,Desarnaud2016a,Felix2018}.

A recent combination of experiments with optical measurements 
and modeling via a thin film model has shown that
when mass supply through the liquid film is insufficient,
growth cannot be maintained in the central part of the 
contact, and a cavity forms in the crystal
within the contact region~\cite{Felix2018}.  In later stages, the 
cavity expands and gives rise to a rim along the edge of the contact.
Such rims have been observed in many previous experiments~\cite{Flatt2007,Roine2012,Li2017,Taber1916}
focusing on the crystallization force produced by the growth
process~\cite{Becker1905,Correns1939,Desarnaud2016a,Naillon2018}, which is known to have important
consequences for deformation and fracturing of rocks,
and the weathering of building materials~\cite{Flatt2002,Espinosa2010}.
However, here wish to focus on the case 
where external forces are small, which correspond for example to 
the experiments of Ref.\cite{Felix2018},
where the crystal was only weakly maintained against the substrate
due to its own weight.

These experiments were also realized with liquid 
film thicknesses in the range from $10$ to $100$nm
due to the presence of nano-scale roughness or dust
between the crystal and the substrate.
Our aim here is to investigate  the possible changes
in this scenario when the thickness of the 
film is decreased down to the nanometer scale
using a thin film model~\cite{Gagliardi2017,Felix2018}
which accounts consistently for thermodynamics, 
non-equilibrium transport processes (diffusion and  advection) 
and crystal-surface interaction.

At the nanoscale, novel ingredients come into play.
The first type of ingredient is related to disjoining pressure effects,
which describe the energetic cost of placing the 
crystal surface at a given distance from the substrate.

The standard theory of disjoining pressure, named the DLVO approach~\cite{Israelachvili1991}, combines 
two effects. The first one is an electrostatic double-layer repulsion
due to the redistribution of charged ions close to the surfaces. 
These forces are exponentially decreasing with the distance.
They are repulsive between similar surfaces but can be 
both repulsive or attractive between dissimilar surfaces~\cite{Israelachvili1991,Overbeek1947}.
The second contribution to the DLVO theory are van der Waals
forces, which lead to power-law interactions between surfaces.
Van der Waals interactions are usually
attractive when a liquid film is present in between the surfaces~\cite{Israelachvili1991,Overbeek1947}. 
In the past decades, significant deviation from the DLVO theory were measured
at short ranges (few nanometers).
These additional (usually repulsive) interactions related, e.g., to the
local ordering or binding of water molecules, are referred to as hydration forces~\cite{Alcantar2003,Delgado2005,Hamilton2010,Diao2016}. 
The sum of power-law attractive forces and of exponential repulsive forces
gives rise to a minimum in the interaction potential, which
corresponds to an equilibrium thickness for the liquid film,
hereafter denoted as $h$~\cite{Israelachvili1991,Alcantar2003}.
This distance is usually in the scale from 1 to 10 nm~\cite{Israelachvili1991}. 
In the presence of such a minimum,
heterogeneous nucleation can occur on the substrate, because
there is a gain of energy when a crystal grows with an interface in this minimum.
Hence, our study could describe growth 
along a flat substrate after heterogeneous nucleation.

In order to account for these effects in our model, we use a disjoining pressure with
an attractive van der Waals contribution together with a generic effective short range repulsion. 
We show that the presence of an attraction makes the appearance of the cavity
discontinuous. Indeed, various quantities, such as the depth
of the cavity, exhibit a jump at the transition.
In addition, there is a minimum
supersaturation needed to induce cavity formation.
However, the non-equilibrium morphology diagram 
describing the occurrence of the cavity 
remains unaffected as compared to the case where 
disjoining pressure is purely repulsive~\cite{Felix2018}.

A second ingredient which becomes relevant
when the film thickness is decreased down to the
nanoscale is viscosity.
Indeed we observe that viscosity hinders
the formation of the cavity. We also show the existence of a critical
viscosity above which cavities
cannot form. We determine the value of the critical viscosity and find it to be proportional to the square
of the film thickness. This result can also be re-formulated
as the existence of a critical thickness
below which the cavity will not form for a given viscosity.

We accompany the presentation of model results
with a semi-quantitative discussion of 
the nano-confined growth of some materials, viz.,
Calcium Carbonate, Sodium Chlorate, Glucose and Sucrose.
Although they belong to disparate classes 
of materials, with time-scales ranging from
second to geological times and contact lengthscales
from microns to centimeters, our modeling 
approach suggests that their behavior
can be globally classified based on a small number of dimensionless 
physical parameters.

\begin{figure}
\center
\includegraphics[width=0.7\linewidth]{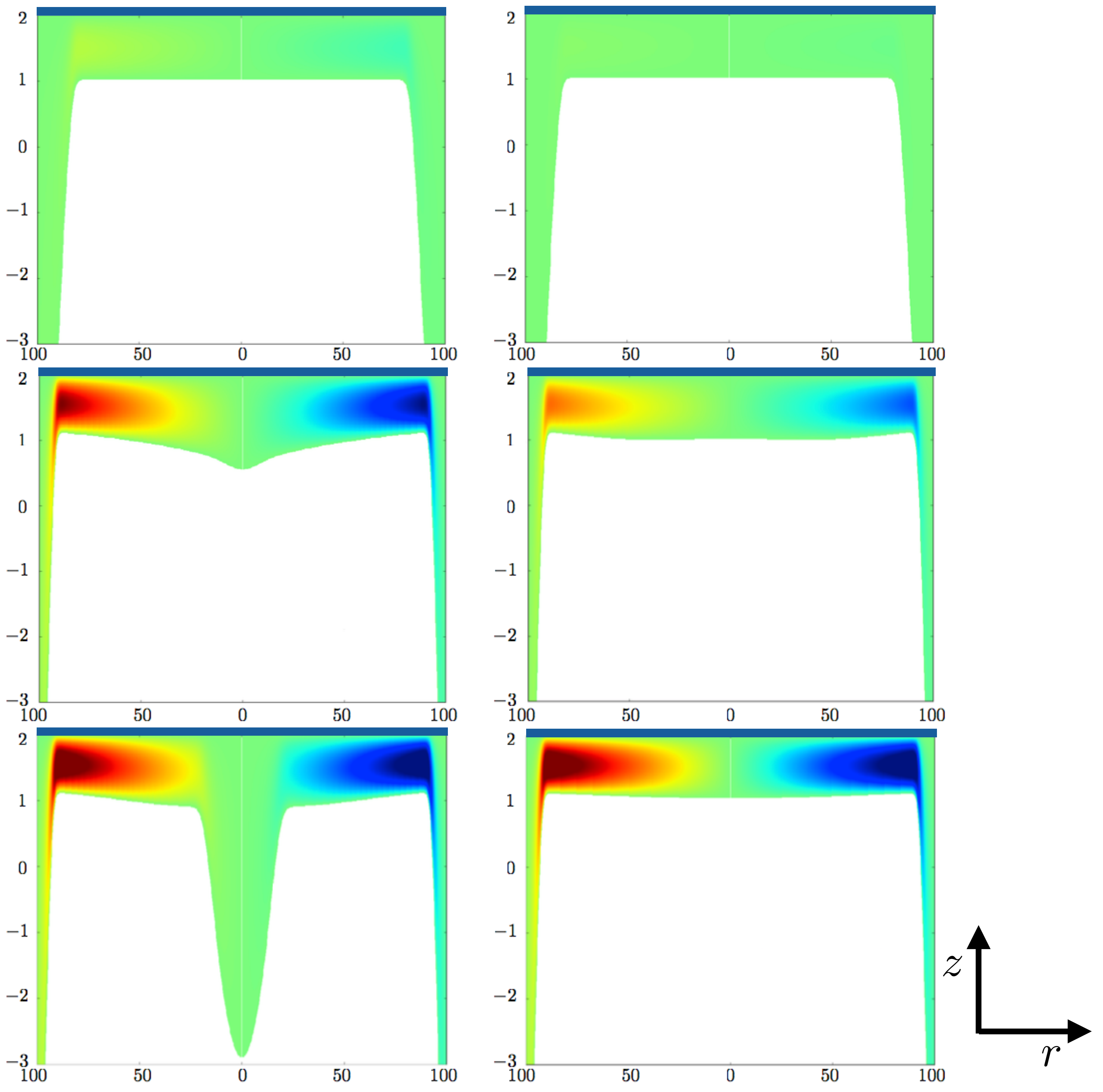}
\caption{Simulation screenshots representing section of an axisymmetric growing crystal (white). Time flows from top to bottom.
The normalized supersaturation is for both panels $\bar{\sigma}_{bc} = 0.21$.  Left column $\bar{\eta} = 10^{-2}$; right column $\bar{\eta} = 10^{-1}$, the cavity is not observed.
% The axis are represented in rescaled units. 
The units of the vertical scale is 1  nm. The substrate is located at $h_s = 2$nm.
The scale of the horizontal axis depends on the material. 
For instance for  NaClO$_3$ the radial scale unit is $3.2$nm.
The color-map represents the liquid velocity in normalized units.
Red color: positive velocities (flow from left to right); 
blue: negative velocities; 
green: vanishing velocity.
The physical liquid velocity depends on the material, for instance
in the left panel for NaClO$_3$ its maximum value (darker color) 
is $u_L \approx 66\mathrm{\mu m/s}$.
\label{fig:screenshots}}
\end{figure}

\section{Model and methods}

We consider a system with a confinement geometry
similar to that of the experiments in~\cite{Felix2018}:
a growing crystal is separated from a flat, impermeable and inert substrate
by a  thin film of solution. 
However, here, the film thickness is assumed to be of the order of nanometers. 
We assume the presence of a macroscopic concentration reservoir 
outside the contact region.

To predict the evolution of the confined interface during crystal growth,
we use the thin film model presented in~\cite{Gagliardi2017}. 
This model describes the growth of a rigid crystal,
and  accounts for diffusion and hydrodynamics in the liquid film. 
We assume that the slope of the crystal surface is small.
Dynamical equations for the interface 
evolution can therefore be obtained by means of the 
standard lubrication expansion~\cite{Oron1997}. 
% {\color{blue}[[Deleted sentence (see original document) because repetitive and probably not very clear for non-theoreticians]]}
Within this limit, due to the slenderness of the film, 
attachment-detachment kinetics is fast as compared to diffusion
along the liquid film. 
This assumption is more robust when considering highly soluble materials.
In addition, we neglect hydrodynamic flow induced by crystal-solution 
density difference, assume the dilute limit and linearized Gibbs-Thomson relation. 
We also assume for simplicity an axisymmetric geometry. 

The system can be visualized in \fref{fig:screenshots},
where the profile of the crystal projected along the radius, 
represented in white, is growing via transport of 
mass from the macroscopic solution reservoir at the boundary of the simulation box
to the crystal surface via the thin film solution.
The velocity field of the liquid is represented 
by the color map and the substrate is represented by the dark-blue rectangle
at the top of the images.

Let us now describe the evolution equations in more details.
Using cylindrical coordinates $z,r$,
the dynamical equation relating the local film thickness $\zeta(r,t)$,
and the vertical rigid-body translational velocity of the crystal $u_z$
along $z$ reads
\begin{eqnarray}
\partial_t \zeta = -B\frac{1}{r}\partial_r\Bigl[r\zeta \partial_r (\Delta\mu/\Omega) \Bigr] - u_{z}\, ,\label{eq:h}\\
\Delta\mu/\Omega
=\tilde{\gamma} \partial_{rr} \zeta +\frac{\tilde{\gamma}}{r}\partial_r \zeta  - U'(\zeta),
\label{eq:mu}
\end{eqnarray}
where $B= D\Omega^2 c_0/(k_BT)$ is an effective mobility,
with  $D$ the diffusion constant,  $\Omega$ the molecular volume,  
$c_0$ the numerical solubility, $k_B$ the Boltzmann constant and $T$
the temperature. In the local chemical potential $\Delta\mu$, 
the first two terms represent the contribution
of surface tension $\gamma(\theta)$ 
($\theta = 0$ surface parallel to substrate). 
These terms are proportional to the surface stiffness 
$\tilde{\gamma} = \gamma(0) + \gamma''(0)$. 
The last term represents the contribution
of the interaction
potential  $U(\zeta)$ between the substrate and the crystal.

Since we here focus on small distances $\zeta$, we need to account for the van der Waals 
contribution to $U(\zeta)$, which is usually attractive
for a liquid film between two solids~\cite{Israelachvili1991}. 
We also included a short range repulsive term 
to account for a generic effective repulsion preventing contact. 
The interaction potential then reads
\begin{equation}
\label{eq:potential}
U(\zeta) = \frac{A}{12\pi}\Bigl( -\frac{1}{\zeta^2} +\frac{2h}{3\zeta^3} \Bigr)\, ,
\end{equation}
where $A$ is the Hammaker constant and $h$ the equilibrium thickness.
It follows that the term appearing in \eref{eq:mu} is 
\begin{equation}
\label{eq:force}
U'(\zeta) = \mathcal{A}\Bigl( \frac{1}{\zeta^3} -\frac{h}{\zeta^4}\Bigr)\, ,
\end{equation}
where $\mathcal{A} = A/6\pi$. 
Given the system under study, in the following we assume $h = 1$nm.

The global balance between viscous forces
produced by hydrodynamic flow and 
the forces resulting from the interaction potential
provides an additional relation which allows one to determine $u_z$:
\begin{equation}
u_{z} \, 2\pi\int_0^R \mathrm{d}r\, r\int_r^R \mathrm{d}r'\, \frac{6\eta r'}{\zeta(r')^3} = 2\pi\int_0^R \mathrm{d}r \, r U'(\zeta)\, .
\label{eq:u}
\end{equation}
Here we have no contribution of external force since we expect gravity
effects to be negligible 
as compared to van der Waals attraction at this scales.
%\Cref{eq:u}, represents the force balance between an external force applied along the z direction, $F_{z}$, the interaction with the substrate and viscous dissipation. 

In practice the dynamical equations were solved in normalized units.
Defining the dimensionless repulsion strength 
$\bar{A} = \mathcal{A}/\tilde{\gamma} h^2$, 
dimensionless variables are the normalized width $\bar{\zeta} = \zeta/h$, 
radius $\bar{r} = r\bar{A}^{1/2}/h$ 
and time $\bar{t} = t B\tilde{\gamma}\bar{A}^2/h^3$.
Rewriting the model equations in a dimensionless form, 
the only parameter explicitly appearing in the equation is the normalized viscosity 
\begin{equation}
\label{eq:viscosity}
\bar{\eta} = \frac{B\eta}{h^2} = \frac{D\Omega^2 c_0}{k_BT h^2} \eta 	\, .
\end{equation}
A large value of $\bar \eta$ indicates a
strong influence of viscosity.
Since $\bar{\eta}\sim h^{-2}$ in \eref{eq:viscosity},
viscosity effects are seen to  be important
when $h$ is small.

The other relevant dimensionless quantities are the normalized system size
\begin{equation}
\bar{R} = \frac{\bar{A}^{1/2} R}{h}\, ,
\end{equation} 
normalized supersaturation
\begin{equation}
\bar{\sigma}=\frac{k_BTh}{\bar{A}\tilde{\gamma}\Omega}\sigma \, , %= \frac{k_BTh^3}{A\Omega}\sigma \, ,
\end{equation}
and the normalized crystal velocity (growth rate)
\begin{equation}
\bar{u}_z = \frac{h^2}{\bar{A}^2\tilde{\gamma}B} u_z\, .
\end{equation}

\begin{table}
\caption{\label{tab:constants}Constants used in the simulations. Other parameters intervening in the scalings are assumed to be independent of the system considered. These are the temperature $T=300$K, the interaction strength $ \mathcal{A} =10^{-21} $J and the typical separation $h=1$nm.
Surface stiffnesses at the crystal water interface are assumed equal to surface tensions and are rough estimations due to lack of data and/or to large variability of it found in the literature.
The last column indicates the solution viscosity at saturation.
}
\begin{indented}
\item[]\begin{tabular}{cccccc}
\br
Material		&	$c_0$	&$\Omega\,[\mathrm{\AA^3}]$&$D\, [10^{-9}\mathrm{m^2/s}]$&$\tilde{\gamma}\,[\mathrm{mJ/m^2}]$ & $\eta\,[\mathrm{mPas}]$	\\
\mr
CaCO$_3\,^{\mathrm a}$		&	$10^{25}$	&	$59$		&	$0.8$ 	&	$100$ & $ 1$\\
NaClO$_3\,^{\mathrm b}$	&	$6\,10^{27}$	&	$69$		&	$0.3$ 	&	$10$& $ 7$\\%exp interpolation + linear and average
Glucose$\,^{\mathrm c}$		&	$3\,10^{27}$	&	$194$	&	$0.2$ 	&	$100$ & $ 10$\\
Sucrose$\,^{\mathrm d}$	&	$3.5 \,10^{27}$	&	$355$	&	$0.2$ 	&	$100$ & $ 100$\\
\br

\end{tabular}
{\footnotesize
\item[]$^{\mathrm a}$ \cite{Li1973,Roine2011,PubChem_calcite} Calcium carbonate is in general characterised by a wide range of solubility due to its strong dependency on carbon dioxide presence. The value in absence of CO$_2$ at $25^{\circ}$ is~\cite{Tegethoff2001} $c_0=0.013\mathrm{g/L}\approx 10^{23}$. However this value can increase of about two orders of magnitude when CO$_2$ is present as is the case in natural environments as sea water~\cite{Miller1952}. We assume the latter.
      \item[]$^{\mathrm b}$ \cite{Seidell1919,Misbah2004,Campbell1966,Lide2005} Data for the diffusion coefficient at saturation was not found.  We estimated this value by extrapolating at higher concentration from~\cite{Campbell1969}. Similarly we extrapolated the data for the viscosity from~\cite{Campbell1966}.
      \item[]$^{\mathrm c}$ \cite{Gladden1953,VietBui2004,PubChem_glucose} There is lack of data for surface tension of glucose-water interfaces. We assume $\tilde{\gamma}\approx 100\mathrm{mJ/m^2}$ as suggested by some experiments on sucrose~\cite{Honig1959}. %The dimensionless viscosity is close to the critical value.
      \item[]$^{\mathrm d}$ \cite{Gladden1953,Honig1959,Mathlouthi1995,PubChem_sucrose}. Diffusion constant was assumed similar to the one of Glucose. %The dimensionless viscosity is far beyond the critical value: cavity appearance should not be possible.
}
\end{indented}
\end{table}

Two sets of simulations with different dimensionless viscosities, 
$\bar{\eta}=10^{-5}$ and $\bar{\eta}=10^{-2}$, were performed. 
They respectively aim at modeling low solubility 
crystals such as Calcium Carbonate (CaCO$_3$), and highly soluble 
crystals like salts and sugars. 
For the latter class, we focused on Sodium Chlorate (NaClO$_3$), 
which was used in our previous work~\cite{Felix2018}, and Glucose. 

The value of the dimensionless viscosity depends 
on the physical parameters as described by \eref{eq:viscosity}.
The values we chose for the simulations are rough estimations. % since the aim of this work is not to give precise quantitative results but rather identify some general behaviors related to concrete physical systems.
For instance Glucose actually lies in an intermediate 
regime between $\bar{\eta} = 10^{-2}$ and $\bar{\eta} = 10^{-1}$.
Some exploratory simulations were also performed at viscosities higher than $10^{-2}$.
Larger viscosities could be encountered in other natural materials as more complex sugars. 
In the case of sucrose for instance,
we have $\eta \approx 100$mPa so that $\bar{\eta} >1$ at saturation~\cite{Mathlouthi1995}.
%while most of the other constants are similar to those of Glucose. 
As a summary, the parameters used in the simulations are listed in \tref{tab:constants}.

Finally, 
% to recover the correct physical units from the numerical solution we also need to know 
the value of the normalized repulsion strength $\bar{A}$ 
is chosen following the same lines as in~\cite{Gagliardi2017}. 
% Since~\cite{Israelachvili1991} $A\approx 10^{-20}$J, 
% we have $\mathcal{A}\approx 10^{-21} J= 1\mathrm{mJ} \cdot h^2 \mathrm{m^{-2}}$.
For simplicity we assume $A\approx 10^{-20}$J~\cite{Israelachvili1991} 
to be the same for all materials considered here.
We then obtain $\bar{A} = \mathcal{A}/\tilde{\gamma} h^2= A/6\pi\tilde{\gamma} h^2$.
In any case, the qualitative behavior is not influenced 
by this parameter which never appears explicitly in the normalized equations,
and only contributes to the spatial and temporal scales
on which phenomena can be observed.

%%%%%%%%%%%%%%%%%%%%%%%%%%%%%%%%%%%%%%%%%%%%%%%%%%%%%%%%%%%%%%%%%%%%%%%%%

\section{Discontinuous transition}

\begin{figure}
\center
	\includegraphics[width=0.7\linewidth]{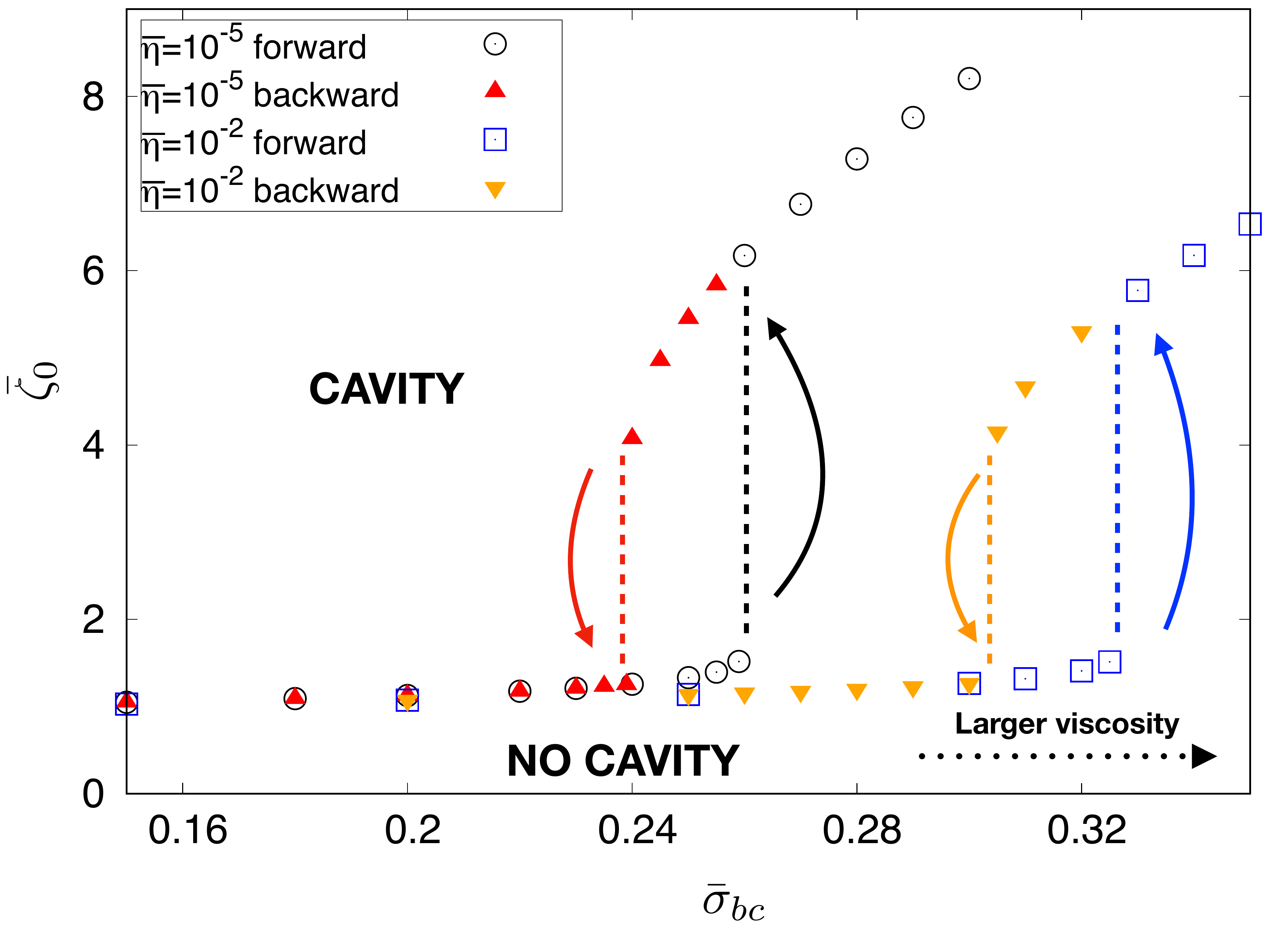}
\caption{Film thickness at the center of the
contact $\bar{\zeta}(r=0) = \bar{\zeta}_0$ 
versus supersaturation $\bar{\sigma}_{bc}$ at the boundary of the simulation box 
at different normalized viscosities $\bar{\eta}$. 
The size of the simulation box is $\bar{R} = 40$. 
% Results are reported in code units.
% To recover physical units prefactors have to be applied. 
The vertical axis is in nanometers. 
The size of the simulation box $R$ and the 
supersaturation scale depend on the material.
Calcium Carbonate, red triangles and black circles: 
$R = 400$nm, $\sigma_{bc} = 0.014 \times \bar{\sigma}_{bc}$; 
Sodium Chlorate, yellow triangles and blue squares: 
$R \approx 127$nm, $\sigma_{bc} = 0.017 \times \bar{\sigma}_{bc} $; 
Glucose, yellow triangles and blue squares: 
$R \approx 400 $nm, $\sigma_{bc} = 0.05\times \bar{\sigma}_{bc}$. \label{fig:hysteresis}}
\end{figure}

We numerically solved \eref{eq:h} and \eref{eq:u} in a circular simulation 
box of fixed radius $R$, and fixed film width $\zeta(R) = \zeta_{bc}$ 
and supersaturation $\sigma(R) = \sigma_{bc}$ 
at the boundary of the integration domain. 
In all simulations we were able to reach a steady state characterized 
by a constant growth rate and crystal interface profile. 
% It should also be remarked that the effective contact 
% radius $L$ is smaller than the total simulation box $R$.
We observe that for low enough viscosities $\bar\eta$, 
a cavity appears when increasing the simulation box radius $R$, 
or the boundary supersaturation $\sigma_{bc}$ . 
In \fref{fig:screenshots} we show two examples of simulations. 
The two columns where realized using different normalized viscosities $\bar{\eta}$,
and keeping the other parameters fixed. 
Simulations at higher viscosity, 
e.g. $\bar{\eta} = 0.1$, do not show the appearance of a cavity. 
% We numerically checked that this does not change with size 
% and supersaturation on a wide range of values.

For the two set of simulations considered, namely $\bar{\eta} = 10^{-2}$ and $\bar{\eta} = 10^{-5}$, we studied the steady state profiles close to the transition.
In \fref{fig:hysteresis} we show as an example the 
variation of the normalized width $\bar{\zeta}(0) = \bar{\zeta}_0$
of the film in the center of the contact as a function of 
the normalized supersaturation $\bar{\sigma}_{bc}$, and
for fixed box size  $\bar{R} = 40$. 
Each dots corresponds to a steady state reached in a single simulation.

Considering a surface which is initially flat and in the minimum 
of the interaction potential ($\bar{\zeta}_0 =  1$), 
and gradually increasing the supersaturation $\bar{\sigma}_{bc}$, 
we observe a sharp jump in the value of $\bar{\zeta}_0$ at the transition. 
This process corresponds to black circles and blue squares
in \fref{fig:hysteresis}.
However if we start with a system beyond the critical supersaturation, thus
featuring a cavity, and slowly decrease the supersaturation $\bar{\sigma}_{bc}$, 
the transition is not observed at the same point, but at a lower supersaturation. 
This is represented by red and yellow triangles in  \fref{fig:hysteresis}. 
Hence, the transition exhibits hysteresis.
A similar behavior is observed when looking at the crystal growth rate. 
This is showed in \fref{fig:hysteresis_vel}, 
where the discontinuity is less apparent especially in the backward transition
(i.e.~when decreasing the supersaturation).

No qualitative difference is observed
between simulations at $\bar{\eta} = 10^{-2}$ and $\bar{\eta} = 10^{-5}$.
The main difference lies in the shift of the transition
towards larger supersaturations when the viscosity is increased.

%%%%%%%%%%%%%%%%%%%%%%%%%%%%%%%%%%%%%%%%%%%%%%%%%%%%%%%%%%%%%%%

\section{Non-equilibrium morphology diagram}

\begin{figure}
\center
\includegraphics[width=0.7\linewidth]{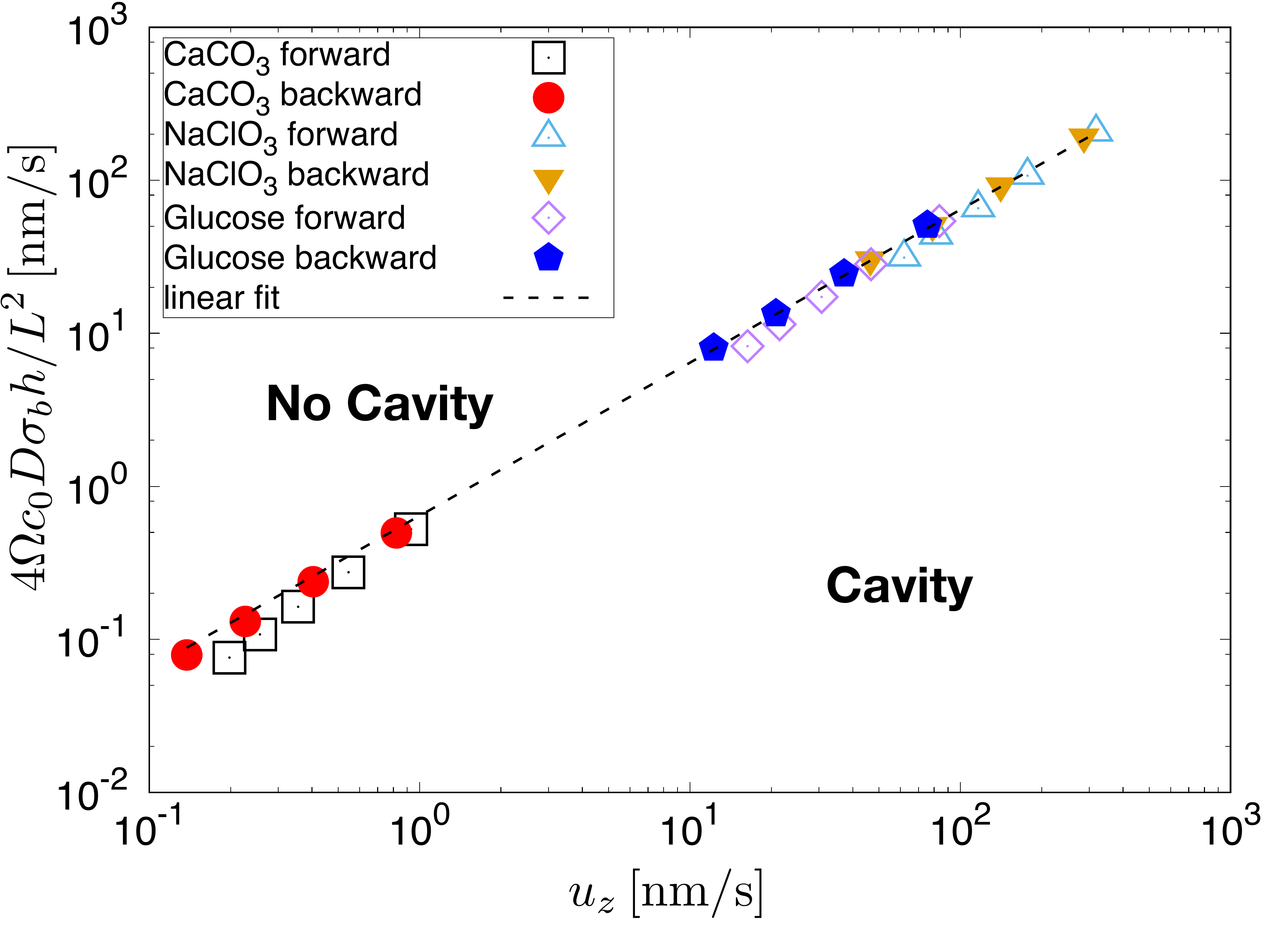}
\caption{ Non-equilibrium phase diagram for cavity formation for different materials and transition pathways.
The scaled viscosity $\bar{\eta}$ is assumed to be $10^{-5}$ for CaCO$_3$ and $10^{-2}$ for NaClO$_3$ and Glucose. 
\label{fig:phase_D}}
\end{figure}

In~\cite{Felix2018}, the conditions under which 
the formation of a cavity can be observed were summarized
in a non-equilibrium morphology diagram.
Let us recall the derivation of the condition for the
transition following the same lines as in~\cite{Felix2018}.
Consider steady state with a flat contact. 
From mass conservation (neglecting the consequences of solute advection),
 the total mass entering the liquid 
film from the boundary of a disc of radius $r$ 
must be equal to the mass entering the crystal, leading to
% [[{\color{green} CHECK SIGNS }{\color{blue} Since I use $J_k$ related to the modulus of the velocity I think is fine}]]
\begin{equation}
\pi r^2J_k=2\pi rhJ_d(r)	\, ,
\end{equation}
where $h$ is the film thickness, 
$J_k$ is the mass flux entering the crystal per unit area 
and $J_d(r)$ is the the diffusion flux entering the liquid film.
Integrating the previous relation and using the identities $J_k = |u_z|/\Omega$ 
where $|u_z|$ is the growth rate, and $J_d(r) = D\partial_r{c}$, 
we obtain the concentration profile $c$.
Then, using the definition of the supersaturation $\sigma=c(r)/c_0-1$, we find
\begin{equation}
\sigma(0) = \sigma_b - \frac{|u_z|}{4hDc_0\Omega}L^2\, ,
\end{equation}
where $L$ and $\sigma_b = \sigma(L)$ 
are respectively the radius 
and the supersaturation at the boundary of the contact area.
Using $\sigma(0)\leq 0$ as condition for cavity formation, 
we obtain  the growth rate at the threshold 
\begin{equation}
\label{eq:Felix}
\alpha |u_z^{cav}| = 4 D\Omega c_0 \sigma_b^{cav}\frac{h}{L^2}  \, .
\end{equation}
Following ~\cite{Felix2018}, 
the heuristic multiplicative constant $\alpha$ is introduced in order
to capture quantitatively the simulation results 
within this simplified  approach.

In order to build a non-equilibrium morphology diagram representing 
the location of the transition (when it exists) in a plane
where the axes are the left hand side and right hand side of \eref{eq:Felix}, 
we need to evaluate the observables $L$ and $\sigma_b^{cav}$.
First, we determine the couple $R$ and $\sigma_{bc}$ at the transition.
Then,  we consider the contact radius $L$ 
from the condition that $\zeta(L)$ 
exceeds the equilibrium position $h$ by $1\%$. 
Finally we obtain $\sigma_b^{cav}$ using
\begin{equation}
\label{eq:sigma}
\sigma_b = \frac{\Delta \mu(L) }{k_B T} = \frac{\Omega}{k_BT}\Bigl [\tilde{\gamma}\kappa(L) - U'(\zeta(L)) \Bigr ]\, ,
\end{equation}
where $\kappa$ is the local mean curvature. 
The procedure is repeated for simulations at different box sizes and viscosities, 
and on the different branches of the hysteresis curve.

The results, shown in \fref{fig:phase_D}, 
confirm the expected linearity of the transition line.
Interestingly, the forward and the backward transitions 
roughly collapse on the same line.
The differences in mass transport kinetics between different materials however 
lead to differences in the orders of magnitude 
of the critical vertical growth velocity $u_z^{cav}$ (from about $0.1$ to $100$nm).
A linear fit for the slope of the transition line leads to $\alpha=0.65\pm 0.04$. 
This result is close to the value $\alpha\approx 0.61$ obtained in~\cite{Felix2018}.
However, the model of~\cite{Felix2018} was different, 
with a purely repulsive potential and a load to maintain the 
crystal close to the substrate. This result suggests
that the constant $\alpha$ could be robust with respect to 
the details of the model.

%-----------------   
%%%%%%%%%%%%%%%%%%%%%%%%%%%%%%%%%%%%%%%%%%%%%%%%%%%%%%%%%%%%%%%

\section{ Critical supersaturation and critical viscosity}

To understand how viscosity can affect the transition 
we resort to a perturbative analysis of the steady-state solution. 
This is done assuming that, just before the transition, 
the profile deviates slightly from the equilibrium configuration
$\zeta = \zeta_{eq} + \delta \zeta$. 
The details of the derivation, reported in~\ref{sec:appendix}, %\sref{sec:appendix},
reveal that the perturbation $\delta\zeta$ exhibits a concave parabolic profile. 
Hence, the thickness $\zeta_0$  in the center 
of the contact increases as the supersaturation increases 
even in the absence of cavity.

This result suggests a simple mechanism for cavity formation.
We use the standard result of the linear stability analysis of an infinite flat profile
of thickness $\zeta$, which indicates that the surface of the crystal should be 
stable when $U''(\zeta)>0$, and unstable when $U''(\zeta)<0$.
This is similar to usual spinodal decomposition~\cite{Cahn1961}.
Hence, the initial profile with $\zeta=h$
is constant and at the minimum of the potential with $U''(h)>0$
corresponds to a stable configuration.
Considering now a non-equilibrium profile with a 
concave parabolic $\zeta(r)$, an approximate criterion for the cavity to form is
that the thickness $\zeta_0=\zeta(r=0)$ at the center of the contact reaches the inflection point $\zeta^{cav}$ 
of the potential, with $U''(\zeta^{cav}) = 0$.
This scenario is consistent with 
a discontinuous transition, since upon destabilization
the thickness $\zeta_0$ in the center of the contact 
becomes larger than $\zeta^{cav}$. Once the instability is initiated, the larger $\zeta_0$,
the larger $U''(\zeta_0)$, and the stronger the destabilization,
leading to a self-amplifying feedback. Note once again
that this behavior is reminiscent of spinodal instabilities~\cite{Mitlin1994,Herminghous1998,Xie1998}.

Using  this simple argument, i.e. $\zeta_0=\zeta^{cav}$, and in the limit of large contacts,
we find an expression for the critical supersaturation:
 \begin{equation}
\label{eq:critical_sup}
\sigma_b^{cav} \approx \frac{\mathcal{A}\Omega}{3k_BT h^3}\Bigl(\frac{1+12\bar{\eta}}{1-12\bar{\eta}}\Bigr)\, .
\end{equation} 
The details of the derivation are reported in~\ref{sec:appendix}.

\begin{figure}
\center
\includegraphics[width=0.7\linewidth]{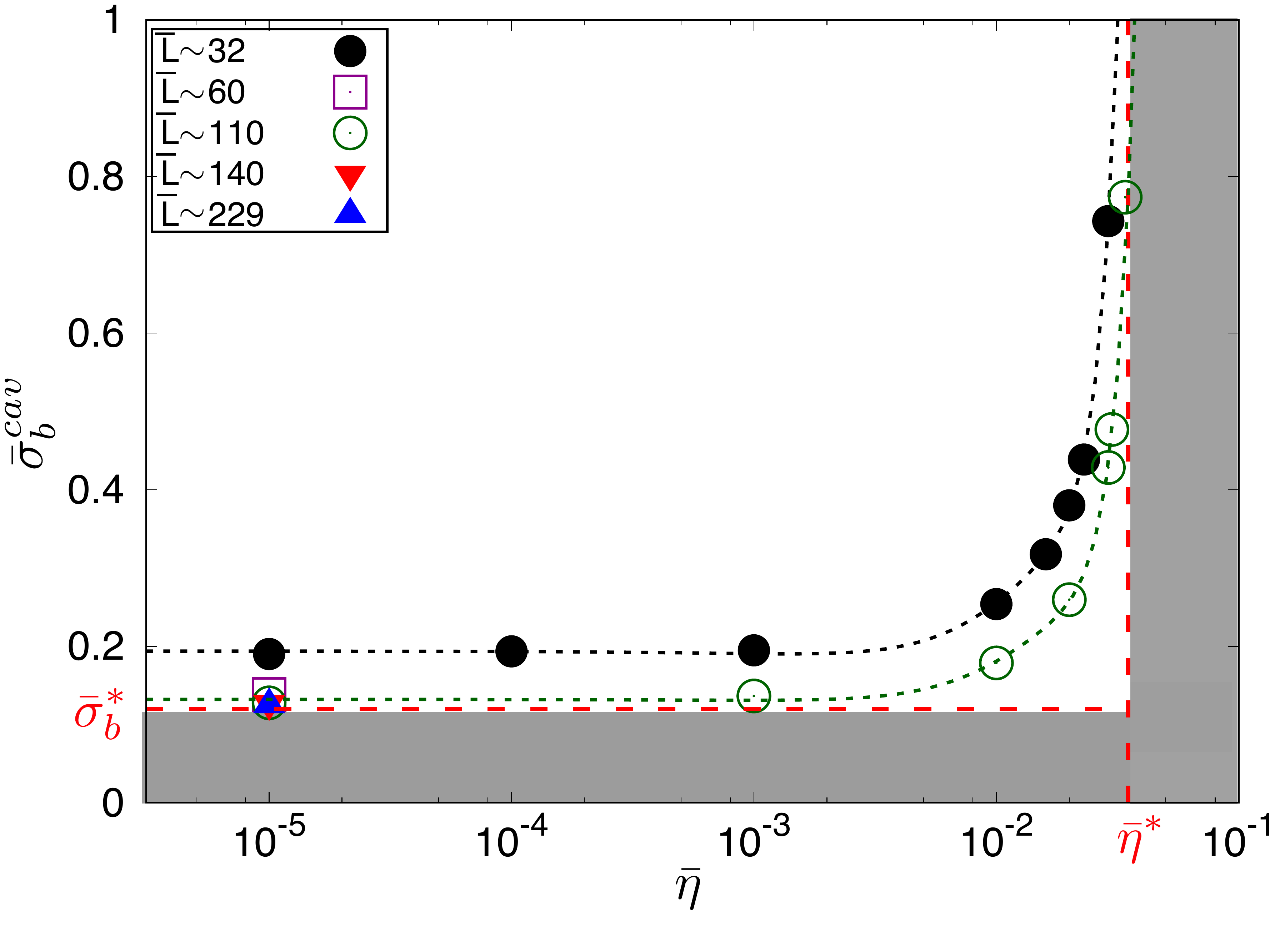}
\caption{Critical supersaturation for the appearance of a cavity 
as a function of viscosity, as obtained from simulations for the forward transition (initially flat contact).
The results are reported in normalized units. 
The critical supersaturation diverges at $\bar{\eta}^*\approx 0.34$. 
For larger normalized viscosities, cavities are not observed in simulations 
independently from the size of the contact (shaded area).
The critical supersaturation converges to a fixed value
when the contact size increase at fixed viscosity, 
as predicted by \eref{eq:crit_sup_full} and \eref{eq:critical_sup}.
At vanishing viscosity the critical supersaturation is $\bar{\sigma}_b^*\approx 0.12$ (red and blue triangles). Cavities cannot be observed independently from the size of the contact below this value (shaded area).
\label{fig:crit_sup}}
\end{figure}

As a first consequence of \eref{eq:critical_sup},
the critical supersaturation $\sigma_b^{cav}$ is expected to
reach a finite value $\sigma_b^*$, when the viscosity vanishes.
This result differs from the behavior of purely repulsive potentials
discussed in \cite{Felix2018}, where vanishingly small supersaturations
were able to destabilize large crystals.
This difference is intuitively understood from the fact that
the supersaturation here needs to  be large enough to lead to 
an escape of the crystal surface from the
potential well at $\zeta=h$. Thus the thermodynamic force related to
supersaturation $\Delta\mu/\Omega$ must be larger
than the disjoining force dragging the interface towards
the minimum of the potential $U'(\zeta^{cav})\approx (\zeta^{cav}-h)U''(h)$.
Since $\sigma_b=\Delta\mu_b/k_BT$, we obtain that $\sigma_b^*=(\zeta^{cav}-h)U''(h)/(\Omega k_BT)$, which is identical to \eref{eq:critical_sup} when $\bar\eta=0$ and $U$ is given
by \eref{eq:potential}. This result, which states
that the the critical supersaturation $\sigma_b^{cav}$ is expected to
reach a constant value when the viscosity vanishes
and the size is large, is confirmed by simulations
in \fref{fig:crit_sup} for small viscosities (blue and red triangles). 
However, the predicted value $\bar{\sigma}_b^*\approx 0.33$ 
is larger than the value observed in simulations 
$\bar{\sigma}_b^{cav}(L\rightarrow \infty) \approx 0.12 $. 
Going back to physical variables $\sigma=\bar\sigma\mathcal{A}\Omega/(k_BTh^3)$,
we find that the critical supersaturation
at vanishing viscosities
is small $\sigma_b^*\sim 10^{-2}$ to $10^{-3}$
for $h\sim 1$nm. Since $\sigma_b^*\sim h^{-3}$, the critical supersaturation decreases quickly
when the equilibrium thickness $h$ increases, and $\sigma_b^*<10^{-5}$ for $h=10$nm. 

The expression \eref{eq:critical_sup} also provides
information about the consequences of viscosity.
For example, it agrees qualitatively with \fref{fig:hysteresis}, 
where higher viscosities were shown to lead to a transition
at higher supersaturations.
In \fref{fig:crit_sup}, we show the normalized critical (forward) supersaturation 
$\bar{\sigma}_b^{cav}$ at different normalized viscosities as obtained by simulations.
This again confirms good qualitative agreement with \eref{eq:critical_sup},
since it agrees both with the increase of $\sigma_b^{cav}$ with 
increasing $\bar\eta$, and with the divergence of  $\sigma_b^{cav}$
for a finite value of $\bar\eta$.
% Furthermore, the approximate expression \eref{eq:critical_sup} 
% exhibits a singularity for $\bar{\eta}^*= 1/12$.
% This is qualitatively consistent with simulations, 
% where we could not observe cavity at high viscosities for any supersaturation.
%  

However, \eref{eq:critical_sup} is quantitatively inaccurate.
For example, the observed threshold at $\bar{\eta}^* \approx 0.034$ 
is lower than the predicted value $\bar{\eta}^*= 1/12\approx 0.08$.
Despite the absence of a quantitatively accurate
expression for the critical supersaturation as a function
of viscosity, it is possible to obtain quantitative insights
about the critical viscosity using the morphology diagram. Indeed,
inserting the parabolic profile $\zeta_{eq}+\delta\zeta$ of the
film in the contact in the force balance equation \eref{eq:u}, 
leads to a second relation valid below the transition
\begin{equation}
\label{eq:vel_visc}
u_z \approx \frac{-4hD\Omega c_0\sigma_b}{(6\bar{\eta} + 1/2) L^2}\, .
\end{equation}
The details of this derivation are presented in~\ref{sec:appendix_2}. 
This expression exhibits quantitative agreement with simulation 
results as illustrated in \fref{fig:visc}. 
%This expression is compared to the numerical results in \cref{fig:viscosity} and proves to be in very good agreement with the simulation results. 
It follows from \eref{eq:vel_visc} that, as viscosity 
increases, the growth rate $u_z$ decreases. 
In addition, for low viscosities the growth 
rate is independent of the viscosity. 
% Since the growth rate has to be larger than a critical value (right side of the transition line in \cref{fig:phase_D}) for the cavity to appear, viscosity can also determine if the formation of the cavity is possible.

Inserting \eref{eq:vel_visc} in \eref{eq:Felix}, 
we find the critical value of the viscosity above which 
the cavity cannot form
\begin{equation}
\label{eq:critical_visc}
\frac{D\Omega^2 c_0}{k_BT h^2} \eta^*=\bar{\eta}^* = \frac{2\alpha-1}{12} \approx 0.025 \pm 0.007\, .
\end{equation}
Interestingly, if we assume the idealized case to hold ($\alpha = 1$), 
we would have obtained $\bar{\eta}^* =1/12$ as in \eref{eq:critical_sup}.
Even though \eref{eq:vel_visc} and \eref{eq:critical_visc} rely on some approximations 
---based on our perturbative analysis and on the heuristic character of the 
parameter $\alpha$--- 
we find that \eref{eq:critical_visc} provides a reasonably accurate prediction close to the value $\bar{\eta}^*\approx 0.034$  
from the full numerical solution of the model.

The discussion of this result can be presented in
two different ways. First, we may assume that 
disjoining pressure effects lead to a fixed
film thickness, assumed for example to be $h\approx 1\,$nm.
Then, using \eref{eq:critical_visc} 
and considering the materials listed in \tref{tab:constants}, 
we find $\eta^* \approx 3.7\times 10^3\,$mPas  for Calcite, 
$\eta^* \approx 12\,$mPas for Sodium Chlorate,
$\eta^* \approx 4.6\,$mPas  for Glucose 
and $\eta^* \approx 1.2\,$mPas  for Sucrose. 
Cavity formation should be hindered or suppressed by viscosity effects
when these values are equal to, or smaller than 
the values of viscosity at saturation reported in the last column of \tref{tab:constants}. These are $1$, $7$, $10$ and $100\,$mPas, respectively. 
Thus, for example we do not expect a cavity to appear for Sucrose
while Calcite could feature a cavity. 
Conclusions on Glucose or Sodium Chlorate are more difficult since the value 
of the critical viscosity is close to the viscosity at saturation.

The threshold can be reformulated in a different manner.
Indeed, since the value of the critical viscosity increases as the square of $h$
there is a critical thickness $h^*$ above which
a cavity can form for a given system.
Using the viscosity at saturation,
we find $h^*\approx 0.016$nm for CaCO$_3$,
$h^*\approx 0.76$nm for NaClO$_3$,
$h^*\approx 1.5$nm for Glucose,
and $h^*\approx 9.2$nm for Sucrose.
These results once again state that cavity formation
should be suppressed for Sucrose with nanoscale confinement.
For other materials with smaller viscosities, the main effect of viscosity
should be to shift the transition as shown in \fref{fig:hysteresis} and \fref{fig:phase_D}.
In general, when the film thickness is larger than $h\approx 10$nm as in~\cite{Felix2018,Li2017},
we expect cavities can form for most materials.

%%%%%%%%%%%%%%%%%%%%%%%%%%%%%%%%%%%%%%%%%%%%%%%%%%%%%%%%%%%%%%%

\section{Discussion}

Some limitations of our approach are discussed in this
section. The first one concerns the difficulty
to analyze strongly anisotropic crystals which exhibit
facets. 
Indeed, the stiffness $\tilde\gamma$ is expected to diverge
at faceted orientations. 
However, in~\cite{Felix2018}, satisfactory quantitative agreement 
with experimental data for faceted crystals was obtained
using a large but finite stiffness. 
Applying this \emph{ad hoc} assumption to
the results of the present paper would not change them
qualitatively. 
However, the value of some physical observables would change. 
If we assume an effective stiffness about $10^3$ - $10^4$ times the 
surface tension~\cite{Felix2018}, 
crystal velocities (see \fref{fig:phase_D}) 
reduce by the same factor. 
In addition, due to our stiffness-dependent normalization
of space variables, our simulations would correspond to larger crystal sizes 
(by a factor $10$ - $100$).
In any case this will not change the measured slope $\alpha$ 
of the non-equilibrium phase diagram nor the value of the critical viscosity 
since these quantities are independent of the stiffness.

A second difficulty is to use continuum
models to describe the consequences of nano-scale 
confinement on diffusion and hydrodynamics.
It is known for example that diffusion
constants in water can vary significantly
with confinement~\cite{Bocquet2010}.
In contrast, the hydrodynamic description of water
with bulk viscosity is known to be quantitatively 
accurate for separations larger than  $\sim 1$~nm~\cite{Bocquet2010}.
At the nanoscale, 
liquids can also be structured  
in the vicinity of solid surfaces. For example,
layering may lead to oscillations
in the disjoining pressure~\cite{Israelachvili1991}.
Additional confinement effects specific to solutions appear
when the liquid film thickness is decreased up to
values that are comparable to the size of the solute molecules.
Such confinement effects could be observed, e.g., for sucrose
which exhibits a molecular size of the order of one nanometer.
Globally, using continuum models to probe nanoscale 
hydrodynamic effects is a challenge. In order to reach
quantitative accuracy, such methods must be based on 
effective models which are calibrated on 
molecular simulations to account for possible deviations from the bulk behavior.
This strategy should allow one to describe some of the consequences of confinement
by means of the thickness-dependence of physical parameters
such as the diffusion constant and the viscosity. Achieving this goal
would be an important step toward the modeling of crystal growth 
with nanoscale confinement.
Indeed, modeling of the growth process in standard molecular dynamics simulations
is difficult due to prohibitive computational time.

Another phenomenon which comes to the fore at the nanoscale is thermal fluctuations.
While the model discussed here is purely deterministic, 
atomistic simulations such as Molecular Dynamics of Monte Carlo Simulations~\cite{Hogberget2016}
 can account for fluctuations.
Thermal fluctuations could trigger the random opening and closure of the cavity 
observed in NaClO$_3$ crystals reported in Ref.~\cite{Felix2018}.
Larger-scale fluctuations or perturbations, such as those due to convection or stirring
in the bulk fluid outside the crystal, should not be relevant here,
since they influence mass transport at scales larger than the thickness of the diffusion
boundary layer  $\ell_{BL}=D/u_L$ at the free surface of the crystal,
which is itself larger than the film thicknesses $h$ considered here.
Indeed, taking $D\sim 10^{-9}\mathrm{m^2/s}$, 
we would need a very large hydrodynamic velocity $u_L\sim 10\mathrm{cm/s}$ 
outside the contact region for $\ell_{BL}$ to reach a scale comparable to 
that of the liquid film in the contact $h\sim 10$nm.

As already mentioned in the introduction,
since it leads to growth perpendicular to the substrate
incorporation of mass in the crystal at contacts may
lead to the generation of forces on the substrate~\cite{Becker1905,Correns1939,Desarnaud2016a,Naillon2018}.
These crystallization forces play an important
role in geology since they are responsible for deformation and fracturing of rocks,
and  are also crucial for the weathering of
building materials~\cite{Flatt2002,Espinosa2010}.
Even though these forces are well characterized at equilibrium via energy
balance~\cite{Steiger2005a,Steiger2005b}, we still lack a precise understanding
of the related non-equilibrium dynamics. A major issue is for instance 
to understand the interplay between the force of crystallization
and the non-equilibrium morphology of the contact~\cite{Flatt2007},
often characterized by the presence of a rim along the edge of the contact 
region~\cite{Flatt2007,Roine2012,Li2017,Taber1916}. 
Despite the absence of external forces in our model,
we hope that our results will provide hints toward
a better understanding of the conditions under which
rims can form.

\section{Conclusions}

In conclusion, we have studied the formation of cavities
in nano-confined crystal surfaces.
Examples are discussed for some model materials 
ranging from poorly soluble minerals (Calcite) 
to high soluble salts (Sodium Chlorate) and sugars. 
% The latter family of materials can induce large 
% viscosities in the liquid film at saturation 
% separating the crystal from the confining substrate.

Cavity formation was recently observed experimentally 
using NaClO$_3$ crystals with liquid film thicknesses that were 
one or two orders of magnitude larger than those used here~\cite{Felix2018}. 
Despite the different scales the resulting non-equilibrium morphology 
diagrams are very similar (with a similar value of the
phenomenological constant $\alpha$). 
This further confirms the robustness 
of cavity formation with respect to variations of physical conditions and materials. 

However, some differences are observed at the nanoscale. 
First, we show that an attractive van der Waals 
interaction induces a discontinuous (subcritical) transition
with hysteresis. 
Moreover, there is a minimum supersaturation
below which cavities cannot form because the driving force is 
not sufficient for the interface to escape from the 
potential well of the disjoining pressure (however its quantitative
value is relatively small when $h$ is larger than $1$nm).
Second, due to the nanoscale width of the liquid film 
separating the crystal and the confining wall, viscosity becomes relevant. 
The effect of viscosity is to shift the transition
toward larger crystal sizes and larger  supersaturations. 
Moreover, the formation of the cavity can also be prevented 
by sufficiently large viscosities. 
 We estimated the relevant critical viscosity above which no cavity should appear.
In practice, such condition could be realized for instance for sucrose.

We hope that our work will inspire novel experimental
investigations or molecular simulations of growth after heterogeneous nucleation
and of growth of sedimented crystals.

\ack
The authors wish to acknowledge funding from the European Union's Horizon 2020 research and innovation program under grant agreement No 642976.

\newpage

\appendix

%%%%%%%%%%%%%%%%%%%%%%%%%%%%%%%%%%%%%%%%%%%%%%%%%%%%%%%%%%

\section{Growth rate as a function of supersaturation}

\begin{figure}
\center
	\includegraphics[width=0.7\linewidth]{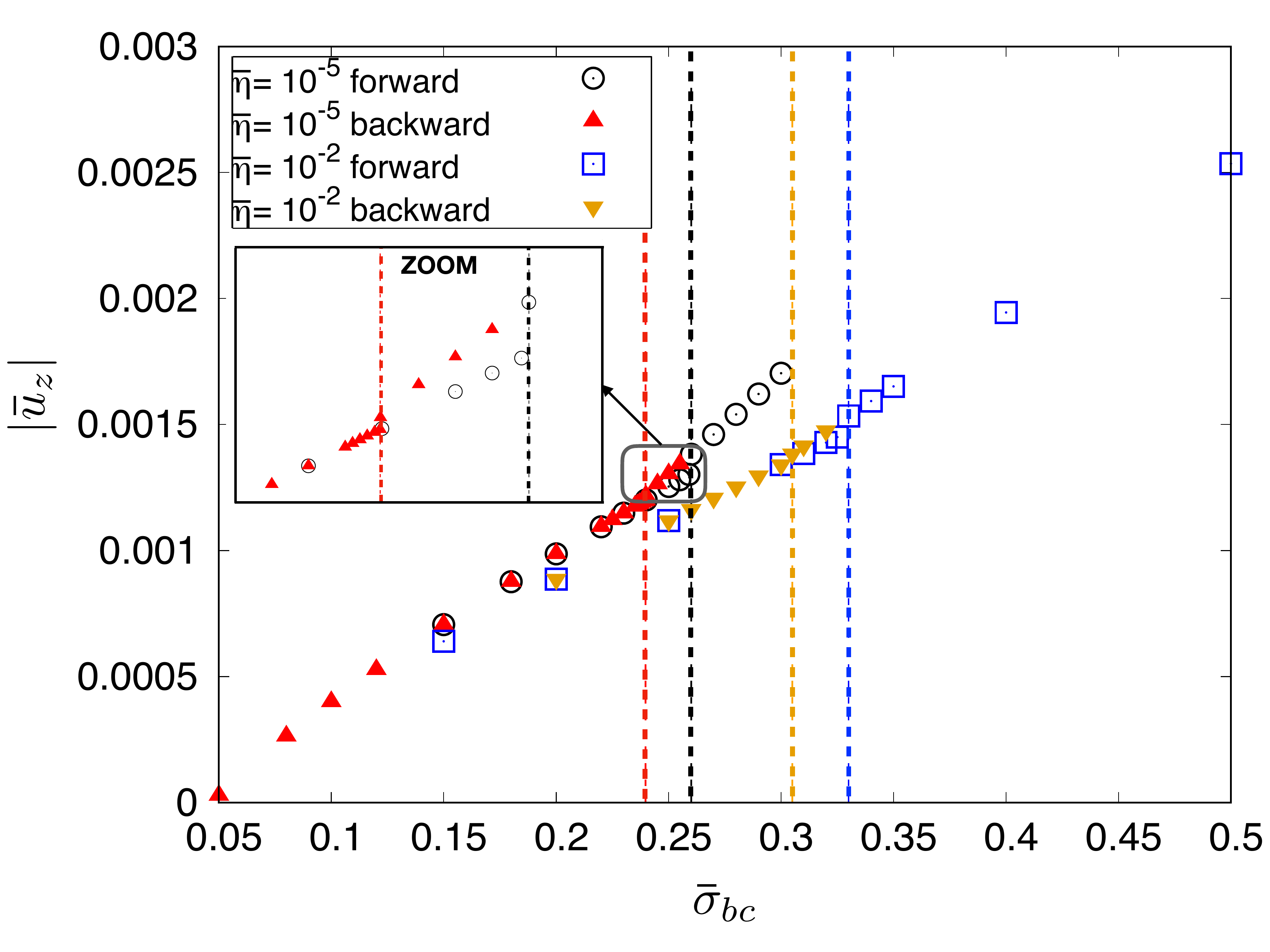}
\caption{Normalized growth rate $|\bar{u}_z|$ versus supersaturation $\bar{\sigma}_{bc}$
at the boundary of the simulation box for different normalized viscosities $\bar{\eta}$. 
The size of the simulation box is $\bar{R} = 40$. 
The system size and scales of the axes depend on the material.
Calcium Carbonate, red triangles and black circles: 
$R = 400$nm, $\sigma_{bc} \approx 0.014 \times \bar{\sigma}_{bc}$, $u_z \approx 6.7\times 10\mathrm{nm/s}\times\bar{u}_z$; 
Sodium Chlorate, yellow triangles and blue squares: $R \approx 127$nm, $\sigma_{bc} \approx 0.017 \times \bar{\sigma}_{bc} $, $u_z \approx 2.1\times 10^5\mathrm{nm/s}\times\bar{u}_z$; 
Glucose, yellow triangles and blue squares: $R \approx 400 $nm, $\sigma_{bc} \approx 0.05\cdot \bar{\sigma}_{bc}$, $u_z \approx 5.5\times 10^4\mathrm{nm/s}\times\bar{u}_z$.
Vertical dashed lines indicate the critical supersaturation at the boundary of the simulation box 
for forward and backward transitions. Their color is the same as
that of the corresponding symbols. \label{fig:hysteresis_vel}}
\end{figure}

In \fref{fig:hysteresis_vel} we show the normalized growth rate $\bar u_z$ 
as a function of the normalized supersaturation at the boundary  of the simulation box
as obtained from numerical solution of \eref{eq:h} and \eref{eq:u}. 
The growth rate responds roughly linearly to 
changes in the supersaturation, and a small jump 
followed by a change of slope is observed at the transition.
Hysteresis is also found here but the discontinuity 
is more apparent when increasing the supersaturation 
from an initial flat surface (forward transition).

%%%%%%%%%%%%%%%%%%%%%%%%%%%%%%%%%%%%%%%%%%%%%%%%%%%%%%%%%%%%%%%%

\section{Perturbation to equilibrium }
\label{sec:appendix}

Using a perturbative approach from the equilibrium solution of \eref{eq:h} and \eref{eq:u},
we here derive approximate expressions for the growth rate and the critical supersaturation. 

As a preamble, we characterize the equilibrium solution itself.
Steady-state solution of \eref{eq:h} and \eref{eq:u} obey
\begin{equation}
\label{eq:steady}
0 = B\frac{1}{r}\partial_r [r \zeta \partial_r (\tilde{\gamma} \partial_{rr}\zeta +\frac{\tilde{\gamma}}{r}\partial_r \zeta -U'(\zeta))] + u_z \, .
\end{equation}
The equilibrium solution is a particular steady-state
equation obeying $u_z = 0$ and
\begin{equation}
\label{eq:chemical_pot}
\tilde{\gamma}\partial_{rr}\zeta_{eq}+\frac{\tilde{\gamma}}{r} \partial_r \zeta_{eq} - U'(\zeta_{eq})=\frac{\Delta \mu_{eq}}{\Omega}\, ,
\end{equation}
where $\Delta \mu_{eq}/\Omega$ is a constant which 
corresponds to the equilibrium chemical potential.
The radius of the contact region is denoted as $L$.
Multiplying \eref{eq:chemical_pot} by $2\pi r$, and integrating 
between the center of the contact at $r=0$ and a radius $r=R>L$,
we find a relation between the equilibrium chemical potential 
and the slope at the boundary of the integration domain
\begin{equation}
\label{eq:chemical_pot_1}
\frac{\Delta \mu_{eq}}{\Omega} = \frac{2\tilde{\gamma}}{R}\partial_r\zeta_{eq}(R) \, ,
\end{equation}
where we have used the relation $2\pi\int_0^R  r dr U'(\zeta)=0$,
corresponding to the equilibrium force balance~\eref{eq:u}.
A second relation can be found when multiplying \eref{eq:chemical_pot} by 
$\partial_r \zeta_{eq}$ and integrating with respect to $r$:
\begin{equation}
\label{eq:contact_angle}
\fl\frac{\tilde{\gamma}}{2} (\partial_r\zeta_{eq}(R))^2 -\Delta U = \frac{\Delta\mu_{eq}}{\Omega} (\zeta_{eq}(R) - \zeta_{eq}(0)) - \tilde{\gamma}\int_0^R \frac{(\partial_r\zeta)^2}{r}\mathrm{d}r \, ,
\end{equation}
where $\Delta U = U(\zeta_{eq}(R)) - U(\zeta_{eq}(0))$. 
%The relation \ref{eq:contact_angle}
\Eref{eq:contact_angle} relating the surface slope $\partial_r\zeta_{eq}(R)$ 
outside the contact to the depth of the potential well $\Delta U$, is
equivalent to a generalized form of the Young contact angle condition.
The integral term in the second equation 
is related to the effect of line tension. 
In the following, we will neglect this term. 

We now assume that the equilibrium profile is flat  $\zeta_{eq}(r)\approx h$ 
with $U'(h)=0$ for $r\leq L$.
Then, we expect $\zeta_{eq}(L) \approx \zeta_{eq}(0) \approx h$,
and combining \eref{eq:contact_angle} and \eref{eq:chemical_pot_1} we find
\begin{equation}
\label{eq:chemical_pot_eq}
\Delta \mu_{eq}  \approx \frac{2\Omega}{L}\sqrt{-2\tilde{\gamma}U(h)}\, ,
\end{equation}
where we assumed that the interaction potential
vanishes far from the contact region $U(\zeta(r>L))\approx 0$.
Note that under these approximations 
the right hand side of \eref{eq:contact_angle} vanishes, 
and this equation is the small 
slope limit of the Young contact angle condition.

\begin{figure}
\center
\includegraphics[width=0.7\linewidth]{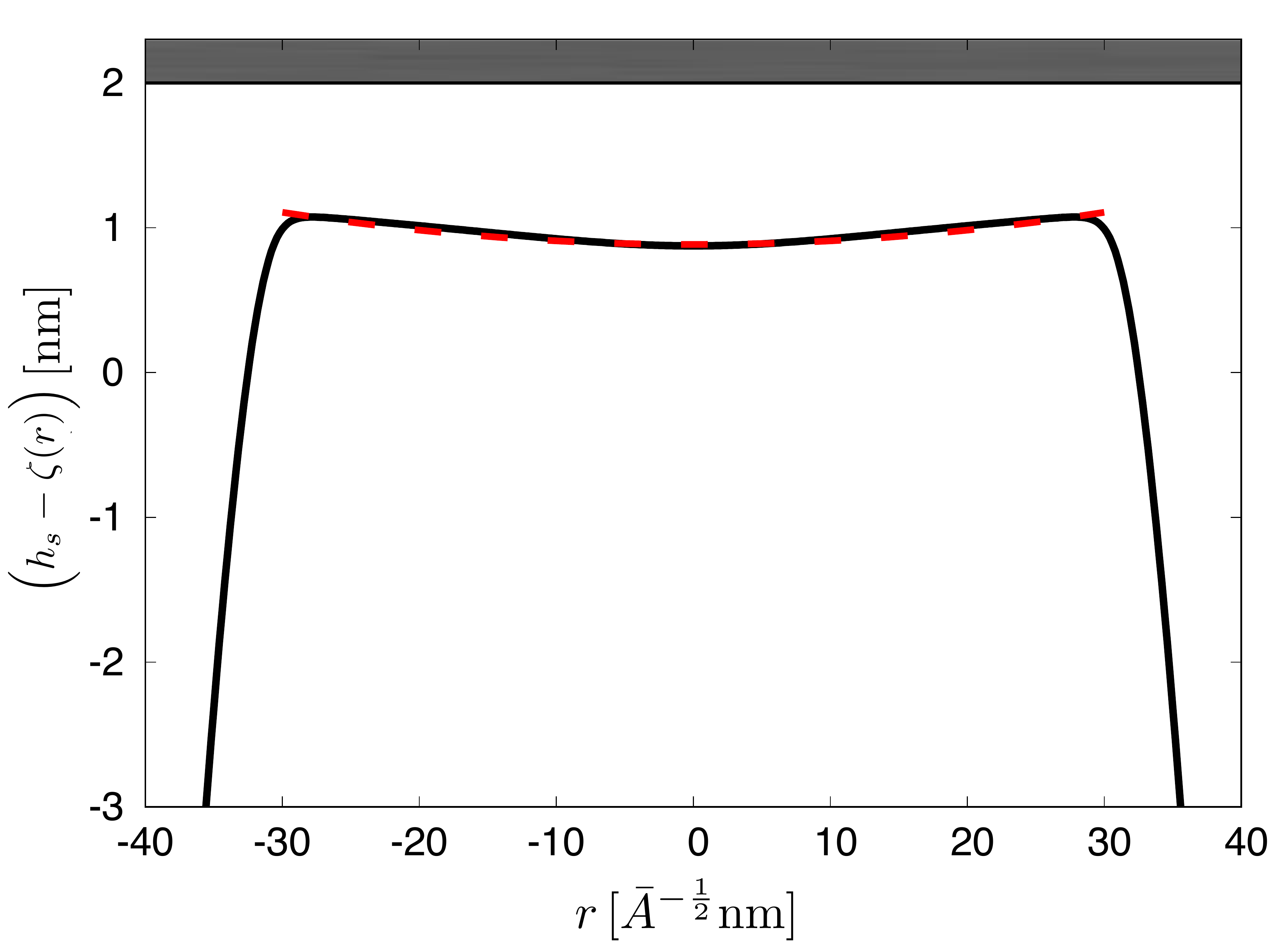}
\caption{Section of the crystal profile  close to the transition. 
The black zone at the top represents the substrate.
The black line is the simulation result. 
The vertical axis is in physical units. 
The horizontal axis scaling depends on the material (via the constant $\bar{A}$). 
Simulation parameters: size of the box $\bar{R} = 40$, 
supersaturation at the boundary of the integration domain $\bar{\sigma}_{bc} = 0.2$.  % Simulation output: contact radius $L=30$, correspondent supersaturation $\bar{\sigma}_b \approx 0.14$, growth rate $\bar{u}_z \approx 10^{-4}$.
The dashed red line is 
% $hs-(h+\delta\zeta)$ 
obtained from \eref{eq:sol} with $L$, $u_z$, $\Delta \mu_b$ 
measured in the simulation.
\label{fig:profile_analytical} }
\end{figure}

Consider now a system below the transition,
so that no cavity is present. The crystal surface
profile is then expected to be close to the equilibrium profile.
We therefore consider the difference $\delta \zeta (r) =\zeta(r) - \zeta_{eq}(r)$ 
between the steady-state solution and the equilibrium solution to be
small. Expanding  \eref{eq:steady} to linear order in $\delta \zeta (r)$,
and integrating two times, we find
\begin{equation}
\label{eq:pert}
\fl\tilde{\gamma} \partial_{rr} \delta \zeta + \frac{\tilde{\gamma}}{r}\partial_r \delta\zeta - \delta\zeta U''(\zeta_{eq}) - \frac{u_z}{2B}\int_r^{L}\mkern-18mu\mathrm{d}r'\frac{r'}{\zeta_{eq}(r')} = \frac{\Delta \mu_b-\Delta \mu_{eq}}{\Omega} ,
\end{equation}
where we have used the 
 parity of $\zeta(r)$ and \eref{eq:chemical_pot},
and we have defined 
the chemical potential at the edge of the contact zone
$\Delta\mu_b = \Delta \mu(L)$ with $\Delta\mu(L)$ given by \eref{eq:mu}.
%\begin{equation}
%\begin{split}
%\frac{\Delta \mu(L)}{\Omega} &=\tilde{\gamma}(\frac{\partial_r\zeta(L)}{r} + \partial_{rr}\zeta(L)) + U'(\zeta(L)) \\
%&\sim \frac{\Delta \mu_{eq}}{\Omega} + \tilde{\gamma}(\frac{\partial_r\delta\zeta(L)}{r} + \partial_{rr}\delta\zeta(L)) + U''(h)\delta\zeta(L). 
%\end{split}
% \end{equation}
Assuming again that in the contact area $r<L$ the equilibrium profile is flat $\zeta_{eq}\approx h$, \eref{eq:pert} can be rewritten as:
\begin{equation}
\fl\tilde{\gamma}\partial_{rr}\delta\zeta + \frac{\tilde{\gamma}}{r}\partial_r\delta \zeta - \delta \zeta U''(h) - \frac{u_z}{4Bh}(L^2 - r^2) = \frac{\Delta\mu_b-\Delta\mu_{eq}}{\Omega}\, .
\end{equation}
A particular solution of this equation is a parabola:
\begin{equation}
\label{eq:sol}
\delta\zeta =\frac{u_z}{4BhU''(h)}(r^2-L^2 + \frac{4\tilde{\gamma}}{U''(h)}) - \frac{\Delta \mu_b-\Delta\mu_{eq}}{\Omega U''(h)}\, .
\end{equation} 
A comparison between this solution and the profile obtained from numerical integration is shown in \fref{fig:profile_analytical} for crystal close to the transition.
The agreement is very satisfactory.

%%%%%%%%%%%%%%%%%%%%%%%%%%%%%%%%%%
\subsection{Viscosity effect on the growth rate}
\label{sec:appendix_1}

\begin{figure}
\center
\includegraphics[width=0.7\linewidth]{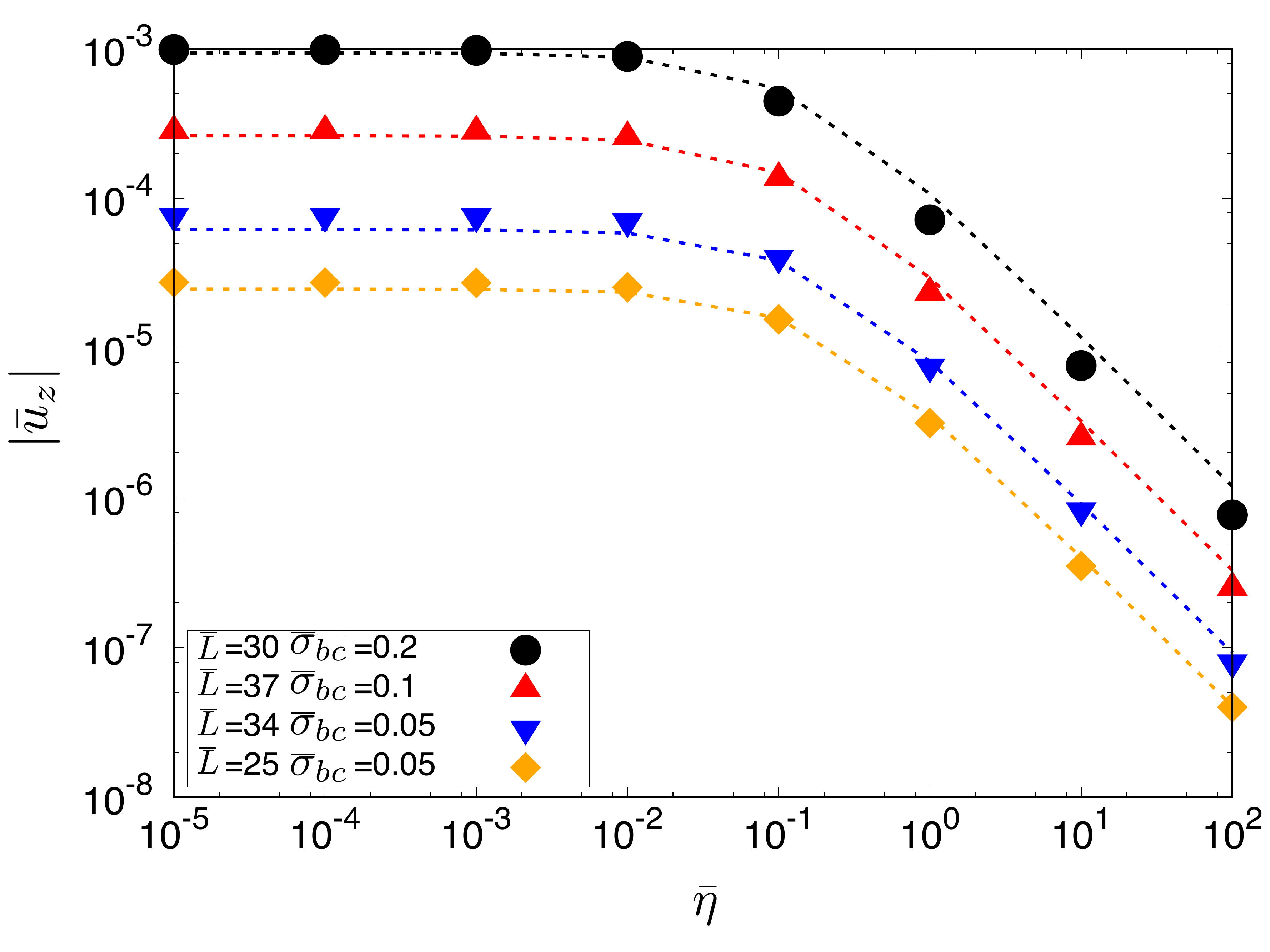}
\caption{Growth rate $|u_z|$ as a function of the viscosity shown in 
normalized units before undergoing the transition (flat growth) for different sizes. The dots are simulation results, the dashed lines were computed using \eref{eq:vel_visc_complete} with $L$ and $\Delta \mu_b(\eta)$ (see \eref{eq:sigma}) measured in simulations and $\Delta\mu_{eq}$ given by \eref{eq:chemical_pot_eq}. The value of the contact size $L$ 
varies weakly when the viscosity is varied.

% In the legend we indicate the boundary supersaturation since σb varies for each dot.
 \label{fig:visc}
}
\end{figure}

Applying a similar procedure to the force balance 
expression in \eref{eq:u} we have to leading order
\begin{equation}
u_z 2\pi \int_0^{L}\mathrm{d}r\, r\int_r^{L}\mathrm{d}r'\, \frac{6\eta r'}{\zeta_{eq}^3(r')} = 2\pi \int_0^{L}\mathrm{d}r\, r \delta \zeta U''(\zeta_{eq}))\, .
\end{equation}
Using \eref{eq:pert} to express the right hand side, we are left with
\begin{equation}
\eqalign{
 \fl u_z 2\pi\int_0^{L}\mathrm{d}r\, r\int_r^{L}\mathrm{d}r'\, r'\Bigl(  \frac{6\eta }{\zeta_{eq}^3(r')} + \frac{1}{2B\zeta_{eq}(r')} \Bigr)=\\ 
= -\pi L^2\frac{\Delta \mu_b - \Delta\mu_{eq}}{\Omega} + 2\pi\tilde{\gamma} L\partial_r\delta\zeta(L)\, .}
\end{equation}
As done previously we assume that in the contact area $r<L$, the equilibrium profile is $\zeta_{eq}\approx h$.
With this assumption the previous relation reduces to
\begin{equation}
\label{eq:vel_intermediate}
u_z \Bigl(  \frac{6\eta}{h^3} + \frac{1}{2B h} \Bigr)\frac{L^4}{4} = -L^2\frac{\Delta \mu_b-\Delta\mu_{eq}}{\Omega} + 2L\tilde{\gamma}\partial_r\delta\zeta(L)\, .
\end{equation}
Using \eref{eq:sol} to express the last term in the right hand size we have
\begin{equation}
\frac{L^4}{4}\Bigl [ (\frac{6\eta}{h^3}+\frac{1}{2Bh}) - \frac{4\tilde{\gamma}}{L^2BhU''(h)}\Bigr ]u_z = - L^2\frac{\Delta \mu_b-\Delta\mu_{eq}}{\Omega}\, .
\end{equation}
%If the radius of the contact region is large $L\gg 2\tilde\gamma^{1/2}/[BhU''(h)]^{1/2}$, 
%we can neglect the second term in the brackets on the left hand side.
We then obtain
\begin{equation}
\label{eq:vel_visc_complete}
u_z = \frac{-4Bh(\Delta\mu_{b}-\Delta\mu_{eq})}{(\frac{6B}{h^2}\eta + \frac{1}{2} -\frac{4\tilde{\gamma}}{L^2U''(h)})L^2\Omega}\, .
\end{equation}
% 
% where we indicated all the implicit dependencies: in general the chemical potential at the boundary of the contact radius depends on the viscosity while given \cref{eq:chemical_pot_eq}, the equilibrium chemical potential depends on the contact radius.
As showed in \fref{fig:visc} the comparison between this relation
and the direct numerical solution of $u_z$ proves to be satisfactory.

Here, we wish to focus on steady-states close to the threshold of cavity formation. 
% Thus, we consider large sizes and supersaturations. 
Since $\Delta\mu_{eq}\sim 1/L$ from \ref{eq:chemical_pot_eq}, 
this term can be neglected far from equilibrium and for large system
sizes where cavity formation occurs. For the same reason we neglect the term of order $1/L^2$.
Finally, assuming the supersaturation is small,
we have $\Delta \mu_b = k_BT\sigma_b$, 
% the explicit expression of $B$ and the definition of the dimensionless viscosity, 
and we obtain \eref{eq:vel_visc}.

\subsection{Viscosity effect on the critical supersaturation}
\label{sec:appendix_2}

% We here by simple arguments show that a threshold viscosity for the existence of the transition at finite supersaturation can be predicted.

% We consider the width $\zeta_0$ at the center of the contact. 
% By a simple mechanical argument we assume that 
% a cavity will appear once $\zeta_0$ surpass the inflection point of the interaction, $U''(\zeta^*)=0$. This corresponds to the point of maximum force attraction towards the minimum of the potential which has a stabilization effect on the interface. It follows that for $\zeta>\zeta^*$ we expect the cavity to initiate.
As discussed in the main text, we expect the cavity to appear
when $\zeta_0>\zeta^{cav}$, where $\zeta_0$ is the width at the center of the contact,
and $\zeta^{cav}$ is defined by the relation $U''(\zeta^{cav})=0$.
Given \eref{eq:force}  and assuming again $\zeta_{eq}\approx h$,
 we find $\zeta^{cav} = 4/3\, h$ and $\delta\zeta^{cav} =\zeta^{cav}-h= h/3$. 
Let us recall \eref{eq:sol} and consider the correction to $\zeta_0$:
\begin{equation}
\delta\zeta(0) = \frac{u_z}{4BhU''(h)}(\frac{4\tilde{\gamma}}{U''(h)}-L^2 ) - \frac{\Delta \mu_b-\Delta\mu_{eq}}{\Omega U''(h)}\, .
\end{equation}
Now we use the condition $\delta\zeta(0) = \delta\zeta^{cav}$ 
for the appearance of the cavity, and deduce the corresponding critical
value of the chemical potential at the boundary:
\begin{equation}
\frac{\Delta \mu_b^{cav}-\Delta\mu_{eq}}{\Omega} 
= \frac{u_z}{4Bh}(\frac{4\tilde{\gamma}}{U''(h)}-L^2) - \delta\zeta^{cav}U''(h)\, .
\end{equation} 
Using \eref{eq:vel_visc_complete} we have
\begin{equation}
\label{eq:crit_sup_full}
\frac{\Delta \mu_b^{cav} - \Delta\mu_{eq}}{\Omega} \approx\frac{\delta \zeta^{cav}U''(h)\Bigl(\frac{6B}{h^2}\eta +\frac{1}{2}-\frac{4\tilde{\gamma}}{L^2U''(h)}\Bigr)}{\frac{1}{2} -\frac{6B}{h^2}\eta }\,.
\end{equation}
Using again the identity $\Delta\mu =k_BT\sigma$, neglecting the last term 
in the denominator ($\sim 1/L^2$) and the equilibrium chemical potential ($\sim1/L$), we obtain \eref{eq:critical_sup}.

\section*{References}

\bibliography{biblio_subcritical}

%\todos
\end{document}